\documentclass[a4paper,11pt]{article}

\pdfoutput=1

\usepackage[english]{babel}
\usepackage[utf8x]{inputenc}
\usepackage[T1]{fontenc}

\usepackage[a4paper,top=3cm,bottom=2cm,left=3cm,right=3cm,marginparwidth=1.75cm]{geometry}

\usepackage{amsmath}
\usepackage{graphicx}
\usepackage[colorinlistoftodos]{todonotes}
\usepackage[colorlinks=true, allcolors=blue]{hyperref}
\usepackage{siunitx}
\DeclareSIUnit{\molar}{M}

\usepackage{textgreek}
\usepackage{pxfonts} 
\usepackage{enumitem}
\usepackage{authblk}
\usepackage{hanging}
\usepackage[many]{tcolorbox}
\newtcolorbox{framefloat}[1][!tb]{arc=0pt,outer arc=0pt,boxrule=0.4pt,
  left=15pt, right=20pt,
  colframe=black,colback=white,float=#1}

\title{Molecular Imprinting: \\
The missing piece in the puzzle of abiogenesis?}
\author[]{K. Eric Drexler}
\affil[]{\small{Future of Humanity Institute, Oxford University \\ eric.drexler@oxfordmartin.ox.ac.uk}}

\begin{document}

\maketitle

\begin{abstract}
In a neglected 2005 paper, Nobel Laureate Paul Lauterbur proposed that \textit{molecular imprinting in amorphous materials}---a phenomenon with an extensive experimental literature---played a key role in abiogenesis. The present paper builds on Lauterbur’s idea to propose \textit{imprint-mediated templating} (IMT), a mechanism for prebiotic peptide replication that could potentially avoid a range of difficulties arising in classic gene-first and metabolism-first  models of abiogenesis. Unlike models that propose prebiotic RNA synthesis, activation, and polymerization based on unknown chemistries, peptide/IMT models are compatible with demonstrably realistic prebiotic chemistries: synthesis of dilute mixtures of racemic amino acids from atmospheric gases, and polymerization of unactivated amino acids on hot, intermittently-wetted surfaces. Starting from a peptide/IMT-based genetics, plausible processes could support the elaboration of genetic and metabolic complexity in an early-Earth environment, both explaining the emergence of homochirality and providing a potential bridge to nucleic acid metabolism. Peptide/IMT models suggest directions for both theoretical and experimental inquiry.
\end{abstract}

\section{Introduction}

Genetic systems enable the open-end accumulation of replicable functional complexity in molecular structures, and the emergence of genetic systems is arguably the pivotal development in the emergence of life. Identifying potential genetic systems compatible with prebiotic chemistries and environments is therefore of considerable interest, yet no generally accepted models have been proposed.

This paper explores the potential implications of \textit{molecular imprinting} and \textit{imprint-directed catalysis} (Section \ref{sec:primer}) for the replication of sequence information in prebiotic peptide polymers, a process which parallels known chemical phenomena and invites experimental investigation (Section \ref{sec:experimental}).\footnote{Note that the present model is compatible with a recent analysis of knowledge and uncertainties regarding the prevalence of life in the universe, an analysis which sketches the IMT model in a footnote. \cite{sandberg2018dissolving}}

\subsection{A proposal for imprint-mediated peptide sequence replication}
\label{sec:proposal}

\begin{figure}[tb]
\begin{center}
\includegraphics[width=0.85\linewidth, trim=0cm 1.5cm 0cm 1.5cm, clip]{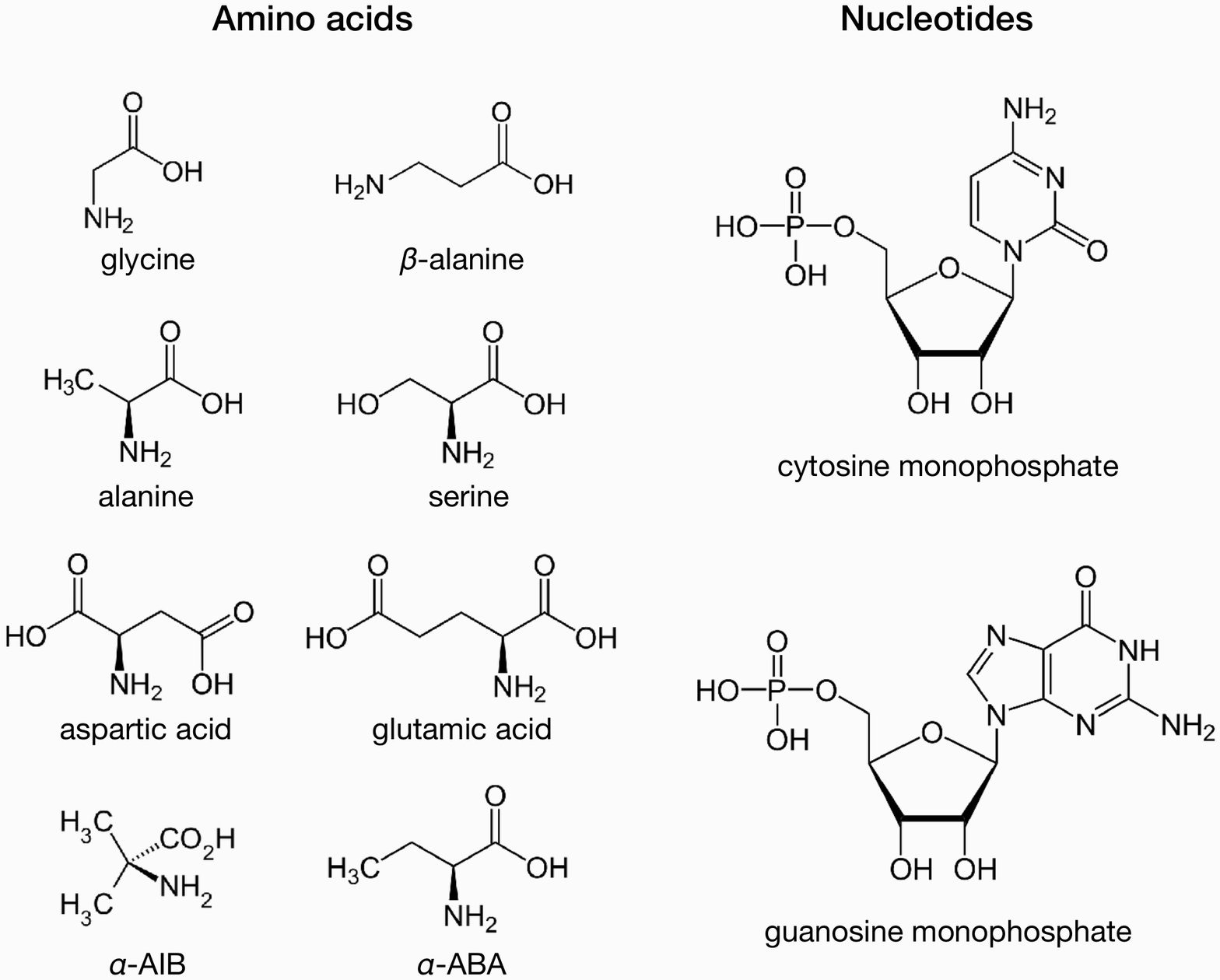}
\caption{The differing structural complexities of nucleotides and amino acids. In contrast to nucleotides, which have been products solely of metabolism or laboratory synthesis, a range of amino acids are produced under prebiotic conditions. The amino acids illustrated above (and others) occur in both extraterrestrial materials and simulated prebiotic reaction mixtures \cite{ring1972prebiotic}, \cite{mccollom2013miller}. Note that many of these, including \textbeta-alanine, \textalpha-AIB, and \textalpha-ABA (above) as well as {\small D} enantiomers of the chiral amino acids, are absent from modern genetically-encoded peptides.}
\label{fig:nucleotides_v_aminos}
\end{center}
\end{figure}

It is widely assumed that the first genetic systems relied on polymers comprising pairwise-complementary (hence necessarily complex) monomers, either modern nucleic acids (\textit{e.g.,} RNA \cite{rich1962problems}) or their functional analogs \cite{schlegel2008duplex}. These models have reached an apparent impasse in which the synthesis of complex genetic polymers requires a complex metabolism, while the evolution of metabolic complexity requires a genetic system \cite{ruiz2013prebiotic}. Realistic prebiotic experiments have both produced (Section \ref{sec:prebiotic_amino}) and polymerized (Section \ref{sec:prebiotic_peptide}) amino acids, yet---perhaps surprisingly to readers outside the field---such experiments have never produced or polymerized nucleic acid monomers. Figure \ref{fig:nucleotides_v_aminos} contrasts the complexity of these classes of monomers.

The present paper proposes a class of models in which replication of information-rich peptide sequences---a genetic process that could potentially enable open-ended evolution---is mediated by \textit{surface molecular imprinting,} a process known to enable robust, strongly selective binding of molecules to imprints formed by (and complementary to) molecules of the same structure and chirality. In this class of models, the consolidation of amorphous materials in contact with imprinting molecules plays the role usually attributed to the templated synthesis of complementary polymers, and could potentially enable the replication of molecules having metabolic functionality without a requirement for homochiral precursors or preexisting metabolic processes.

Evolutionary processes mediated by molecular imprinting and peptide synthesis could thus potentially avoid long-standing difficulties with models of abiogenesis in which nucleic acids are primary (Section \ref{sec:long-standing_questions}), while establishing a context in which nucleic acids could more plausibly arise. The concept of imprint-mediated genetics suggests novel lines of theoretical and experimental inquiry (Section \ref{sec:experimental_and_computational}).

\subsection{Lauterbur’s proposal for imprint-based metabolism} 

Lauterbur (\cite{lauterbur2005demystifying}, \cite{lauterbur2008spontaneous}), introduces the fundamental idea of imprint-mediated catalysis as a potential mechanism for molecular replication in a prebiotic environment. In this paper, Lauterbur proposes that imprinting could support self-sustaining reaction networks,\footnote{ “\ldots a replication reaction, in which \textit{n} molecules can be involved in making \textit{m} imprints, and \textit{m} imprints can catalyze \textit{p} reactions [such that] the products of those reactions are just the molecules responsible for the original imprint functions\ldots. Such a cycle depends on all the relative rates and equilibrium constants being within certain limits\ldots”} but does not examine the potential for imprint-mediated transmission of sequence information and consequent evolution of genetic and functional complexity. Lauterbur’s pioneering concept is thus  uniquely related to, yet quite different from, proposals for peptide/IMT genetic processes.

\subsection{Content outline:}

The argument presented here has multiple facets, reaching from a review of molecular imprinting, to characteristics of anticipated prebiotic environments, to potential mechanisms for the emergence of genetic and metabolic systems, to comparisons with proposals for nucleic-acid based abiogenesis. Accordingly, it may be useful to provide a brief topical outline:

\begin{itemize}[itemsep=0pt]
\item Section \ref{sec:primer} reviews the fundamentals of molecular imprinting and imprint-directed catalysis, establishing the basis for IMT’s relaxed requirements for monomer complexity, chirality, and concentrations.
\item Section \ref{sec:candidate_mechanism} compares and contrasts IMT processes to the monomer-level, polymer-mediated templating processes that underlie nucleic acid replication.
\item Section \ref{sec:prebiotic} reviews current knowledge of prebiotic chemistries and conditions, including the availability of  amino acids, mechanisms for peptide polymerization, and the availability of potential imprinting media.
\item Section \ref{sec:toward_peptide_replication} examines prospects for peptide sequence replication through product-directed, imprint-mediated chain extension and ligation.
\item Section \ref{sec:fidelity} considers copying fidelity and threshold requirements for supporting Darwinian selection and cumulative evolutionary change. \textit{(This section completes the discussion of the basic peptide/IMT model of abiogenesis).}
\item Section \ref{sec:metabolic_evolution} explores potential paths from surface-bound peptide/IMT systems to self-replicating systems with structures and metabolic complexity comparable to cells, seeking paths that avoid requirements for implausible mechanisms or discontinuous change.
\item Section \ref{sec:experimental_and_computational} outlines experimental approaches to studying potential peptide/IMT chem\-istries and outlines computational approaches to studying IMT system dynamics and evolutionary capacity.
\item Section \ref{sec:long-standing_questions} describes longstanding puzzles in models of abiogenesis and how those puzzles can apparently be resolved within peptide/IMT models.
\item Section \ref{sec:conclusions} provides a concluding summary.
\end{itemize}

\begin{framefloat}[tb]
\medskip
{\large\textbf{Key terms and concepts:}\par}
 \begin{hangparas}{.15in}{1}
\medskip
\textbf{\textit{Molecular imprint.}} A structure induced by contact between an \textit{imprinting} or \textit{templating} molecule and a labile, disordered medium that is subsequently consolidated, \textit{e.g.,} by cross-linking. Imprints are cavities that retain molecular-level complementarity to imprinting molecules after their exit.

\smallskip
\textbf{\textit{Polymer mediated templating (PMT).}} The familiar process in which a polymer templates a \textit{complementary polymer} that in turn templates the formation of a polymer identical to the first.

\smallskip
\textbf{\textit{Imprint-mediated templating (IMT).}}   A proposed process in which a polymer templates a \textit{complementary imprint} that in turn templates the formation of a polymer identical to the first. 

\smallskip
\textbf{\textit{Genotype.}}  The \textit{information} that specifies the structure of a replicable molecule, \textit{e.g.,} its monomer sequence. 

\smallskip
\textbf{\textit{Direct phenotype.}}  The \textit{physical structure} of a replicable molecule implied by its genotype. 

\smallskip
\textbf{\textit{Indirect phenotype.}}  A replicable molecule may induce \textit{external structures and processes} that influence its own replicative success; these structures and processes constitute its \textit{indirect phenotype.}
\medskip
 \end{hangparas}
\end{framefloat}

\section{Molecular imprinting: A primer}
\label{sec:primer}

Molecular imprinting occurs under remarkably general conditions and operates on a wide range of molecules and materials, yet informal discussions suggest that the phenomenon will be unfamiliar to most readers. This section briefly surveys the scope of molecular imprinting (mechanisms, properties, and applications), and introduces the key phenomenon of structure-specific, imprint-directed catalysis.

\subsection{Molecular imprinting been widely studied and applied}

In molecular imprinting \cite{komiyama2003molecular}, \cite{whitcombe2014molecular}, a labile imprinting medium (\textit{e.g.,} a mixture of unpolymerized or incompletely-polymerized monomers) reorganizes around an imprinting molecule (here, think of solvation), forming  energetically favorable non-covalent contacts; after consolidation of the imprinting medium (\textit{e.g.,} by polymerization and cross-linking to form a solid or gel) departure of the imprinting molecule leaves a complementary cavity that will selectively bind molecules of the same kind. Since its discovery in the 1980s \cite{arshady1981synthesis}, non-covalent molecular imprinting has been widely studied and applied, particularly in chemical sensing and molecular separation \cite{chen2011recent}, \cite{hussain2015molecular}, \cite{chen2016molecular}.

Molecular imprinting routinely yields structures that bind molecules with antibody-like specificity \cite{wulff1995molecular}, \cite{hoshino2008peptide}, \cite{ye2008molecular}. Imprinting is perhaps unique in operating on mixtures of achiral precursors to produce structures that, although disordered, interact with a specificity that rivals that of biological macromolecules.

\subsection{Diverse media can form molecular imprints}
\label{sec:diverse_media}

Molecular imprinting is a general phenomenon that has been demonstrated in a host of media, not only polymerizable fluids, but also molecular monolayers \cite{balamurugan2011molecular}, thicker films of surface-deposited molecules \cite{shi1999template}, \cite{rick2006using}, micellar structures \cite{awino2017sequence}, \cite{fa2017peptide}, amorphous silica \cite{katz2000molecular}, \cite{markowitz2000catalytic}, and mixed organic/inorganic materials \cite{diaz2005molecular}, \cite{hu2010recognition}.

\begin{table}[tb]
\begin{center}
\caption{Molecular interactions that can occur among prebiotic molecules and contribute to binding specificity in both molecular imprinting and biomolecular systems.}
\medskip
\medskip
\includegraphics[width=\linewidth, trim=1.4cm 4cm 1.4cm 4cm, clip]{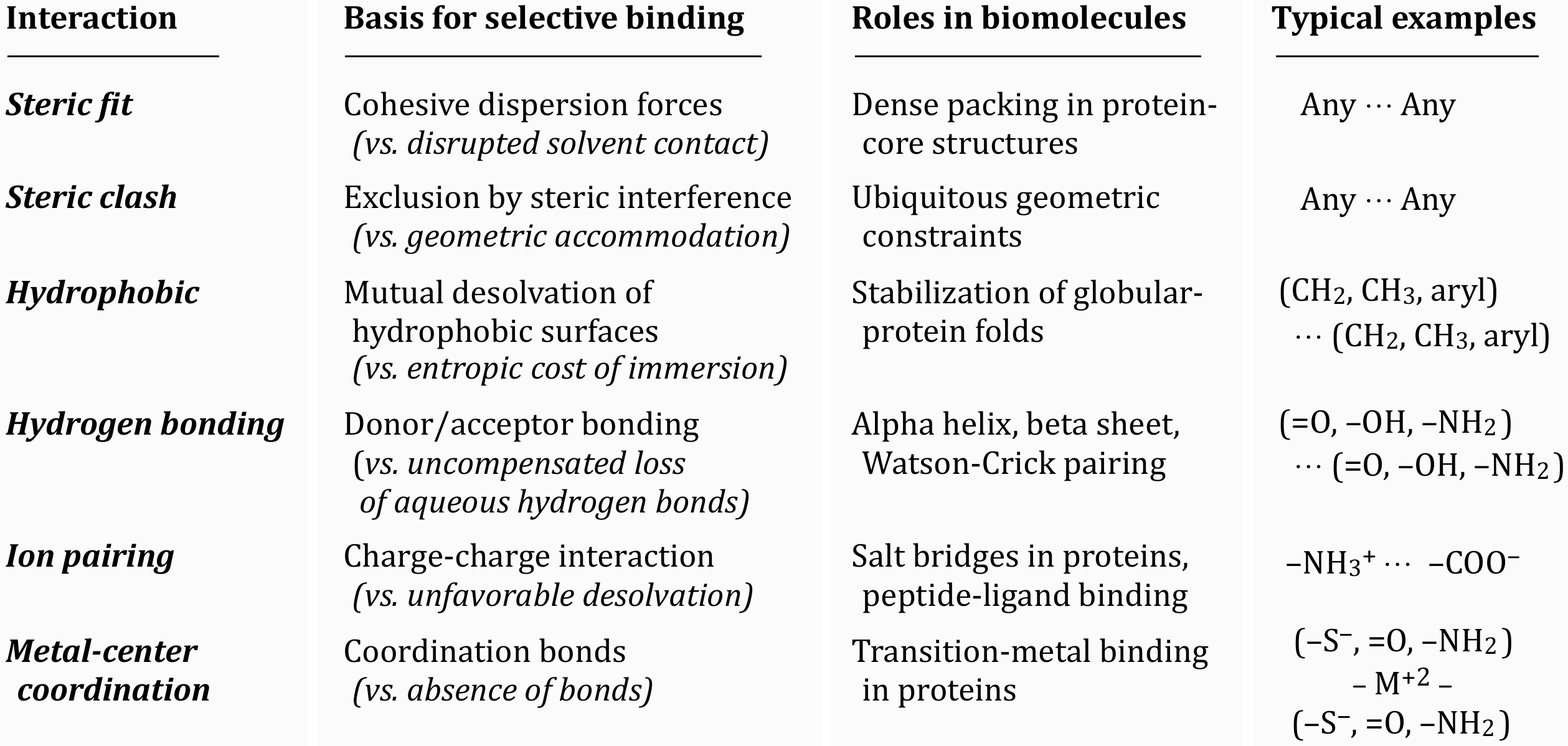}
\label{tab:interactions}
\end{center}
\end{table}

The key features of imprinting media are initial physical lability and subsequent consolidation to a stable amorphous structure. The generality of imprinting processes and media virtually guarantees that a range of imprintable media can be found among substances produced in prebiotic chemical processes \cite{lauterbur2005demystifying}, \cite{lauterbur2008spontaneous}; see Table \ref{tab:interactions}.

In a consolidated imprinted medium, structural memory is a requirement, but rigidity is not---indeed, many imprinting systems employ gels in which molecular porosity and flexibility allow induced fit, and even diffusion of small molecules to and from imprints formed beneath the surface. However, to bind and release larger molecules, imprints (termed “surface imprints”) must have some exposure to the external solvent. In the context of potential imprint-mediated genetic processes, “imprinting” refers to surface imprinting in this sense.

\subsection{Imprints can operate as selective catalysts}

Like antibodies \cite{jacobsen1994antibody}, molecular imprints can act as catalysts \cite{ramstrom1999synthesis}. Common strategies for inducing imprint-mediated catalysis imprint a medium with a molecular transition-state analog \cite{wulff2011design}; because the imprinting molecule differs from the product, however, transition-state imprinting is not directly applicable to molecular replication. For this, we require \textit{product-directed catalysis} induced by imprinting with the target reaction product.

In the peptide domain, it has been found that imprints templated on disulfide-linked cyclic peptides can catalyze corresponding cyclization of the linear peptide sequence \cite{cenci2016guided}, \cite{shen2016catalytic}. General principles of molecular interaction and catalysis suggest that product-directed imprint catalysis can likewise effect chain extension and ligation by peptide bond formation (Section \ref{sec:peptide_ligation}), but questions regarding the viability and generality of this mechanism call for experimental investigation (Section \ref{sec:experimental}).

With these remarks as a caveat, Section \ref{sec:candidate_mechanism} explores the potential role of peptide/IMT processes as a genetic mechanism.

\begin{figure}[thb]
\begin{center}
\includegraphics[width=\linewidth, trim=1cm 7cm 1cm 7cm, clip]{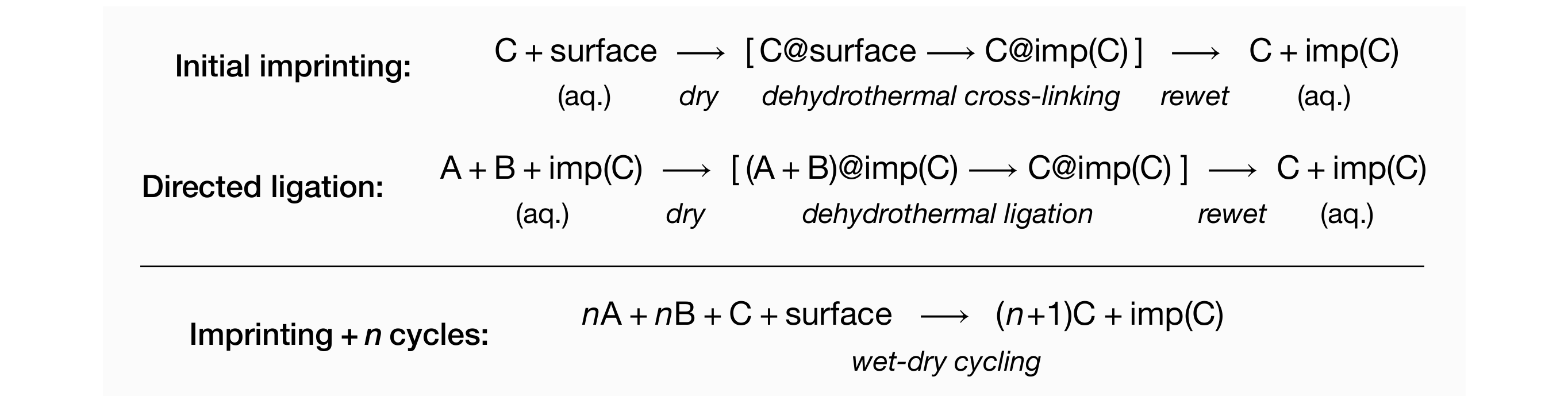}
\caption{Schematic peptide/IMT steps. \textit{A} and \textit{B} represent subsequences of \textit{C}.}
\label{fig:imprinting_cycles}
\end{center}
\end{figure}

\section{IMT as a candidate genetic mechanism}
\label{sec:candidate_mechanism}

Proposed IMT processes parallel yet contrast with \textit{polymer-mediated templating} (PMT), the basis for nucleic acid replication. In brief, IMT entails the formation of an imprint complementary to a polymer sequence (here, a peptide), followed by binding and ligation of contiguous subsequences of that sequence to produce copies of the original (Figure \ref{fig:imprinting_cycles}).\footnote{Note that this is a simplified story: Section \ref{sec:nonlinear} discusses the role of more general operations and structures, while for convenience in discussion, “ligation” will typically be taken to include the addition or dimerization of amino acid monomers, sequences of length 1.} Nucleic-acid PMT is a familiar concept and will be taken as a point of reference in examining the potential role of peptide IMT as a genetic mechanism in high-entropy prebiotic environments.

\begin{figure}[tb]
\begin{center}
\includegraphics[width=0.9\linewidth, trim=0.5cm 2cm -0.5cm 2cm, clip]{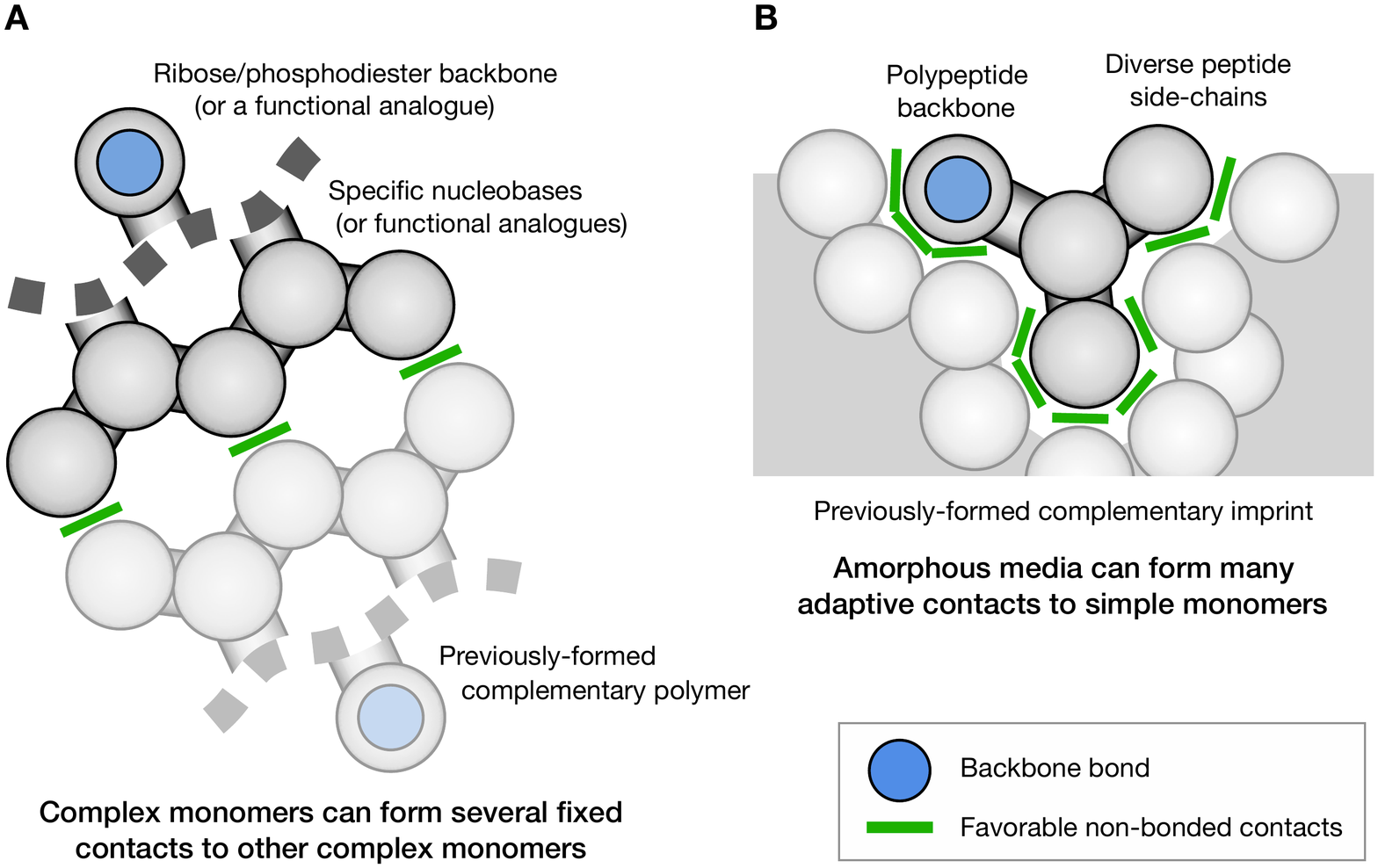}
\caption{Schematic diagrams of contrasting contact geometries in polymer- and imprint-mediated templating. In polymer-mediated templating (\textbf{A}) a small repertoire of complex, distinctive monomers must display pre-existing, pairwise-complementary sets of features, here loosely modeled on the interface of a G:C base pair. Imprint-mediated templating (\textbf{B}) would instead exploit induced, solvation-like complementarity between monomers and an initially-labile amorphous medium. As illustrated, imprints can partially surround small structures with complementary contacts (of diverse kinds, Table \ref{tab:interactions}), and as a consequence, imprints commonly exhibit antibody-like binding specificity \cite{wulff1995molecular}, \cite{ye2008molecular}.}
\label{fig:nucleotide_v_imprint}
\end{center}
\end{figure}

\subsection{Imprint-mediated \textit{vs.} polymer-mediated templating}

In nucleic-acid PMT, a polymer strand templates a second polymer strand that subsequently templates a strand identical to the first; in proposed peptide IMT, a polymer strand templates an imprint that subsequently templates a molecule identical to the first (Figure \ref{fig:nucleotide_v_imprint}). Thus, both IMT and PMT processes require two stages, but the operations differ:

\begin{itemize}[itemsep=0pt]

\item In the first stage, PMT requires the formation of particular bonds between specific, complementary monomers assembled from solution, while IMT requires only non-specific cross-linking a pre-existing dense, heterogeneous medium.

\item In the second stage, IMT selects peptides and monomers from solution through antibody-like interactions (partially surrounding the bound molecule), while in PMT, geometric constraints on monomer pairing preclude similar interactions, even if monomers are large and complex (see Figure \ref{fig:nucleotide_v_imprint}).

\end{itemize}

Known PMT mechanisms require a pre-existing catalyst (a polymerase) in both stages, yet in IMT, imprint formation (Stage 1) requires no catalysis, while in ligation and chain extension (Stage 2), the imprint itself acts as a catalyst. Accordingly, IMT could potentially replicate information-rich polymers that in themselves perform neither molecular recognition nor catalysis, while PMT calls for simultaneous emergence of molecules that perform both functions.

\subsection{High entropy and simple molecules are compatible with IMT}

To enable specific pairing, PMT-based processes require monomers drawn from what must in practice be a small repertoire of physically distinctive complementary structures (in modern biology, only 4). In IMT-based processes, by contrast, antibody-like binding can select small molecules from high-entropy mixtures, providing scope for simple, diverse monomers with chemical functionality unrelated to replication. These properties of IMT processes avoid implicit requirements for pre-existing metabolism while (in peptide IMT) constituting a pre-adaptation for the later emergence of metabolic processes from prebiotic environments.

\section{Prebiotic chemistries: \\Amino acids, peptides, imprinting, and chirality}
\label{sec:prebiotic}

This section briefly reviews current knowledge of prebiotic chemistries and conditions from the perspective of proposed peptide/IMT processes. In brief, known or expected prebiotic conditions can give rise to a range of monomers and polymers; among these are amino acids, peptides, and potential imprinting media.

\subsection{Known and model prebiotic processes produce amino acids}
\label{sec:prebiotic_amino}

In a pioneering experiment, Miller and Urey \cite{miller1953production} demonstrated the synthesis of amino acids from reducing gas mixtures exposed to electrical discharge, an experiment intended to simulate reactions under plausible early-Earth conditions. Subsequent studies of diverse models of prebiotic environments have demonstrated the synthesis of amino acids (including a range of biological amino acids---Ala, Asp, Glu, Gly, Ile, Leu, Pro, Ser, Thr, and Val \cite{longo2013simplified}) as well as other  amine and carboxylic acid compounds of potential relevance to peptide/IMT processes \cite{ring1972prebiotic}, \cite{cleaves2008reassessment}, \cite{johnson2008miller}, \cite{parker2011primordial}, \cite{mccollom2013miller}, \cite{cleaves2014amino}; typical reaction products also include an abundance of insoluble organic polymers \cite{wollrab2016chemical}. Soluble materials from carbonaceous chondrite meteorites are broadly similar, containing substantial concentrations of diverse amino and imino acids, together with dicarboxylic acids, amines, and diamines \cite{pizzarello2006chemistry}, \cite{aerts2016contamination},\cite{ring1972prebiotic}; these again occur in conjunction with insoluble organic polymers.

\subsection{Peptide polymerization can occur on hot, dry surfaces}
\label{sec:prebiotic_peptide}

The synthesis of short peptide polymers (oligomers) has likewise been demonstrated under prebiotic conditions \cite{rode1999combination}. Proposed and demonstrated prebiotic syntheses of peptides have commonly employed wet-dry cycles at moderately elevated temperatures (dehydrothermal cycling), first demonstrated by Lahav \cite{lahav1978peptide} and applied in a series of more recent studies. Under these conditions, mineral surfaces \cite{lambert2008adsorption} and salt evaporites \cite{rode1999combination} can catalyse peptide synthesis.

Dehydrothermal peptide formation been observed at temperatures ranging from 80 to 160°C \cite{rode1999combination}. Temperatures on the early Earth spanned this range \cite{sleep2010hadean}, initially far higher worldwide, then gradually falling to moderate temperatures in environments that would nonetheless have included volcanic hot spots. Through the action of tides, waves, and intermittent precipitation, the early Earth would have exposed large areas (trillions of square centimeters?) of mineral surfaces (clays, sands, rocks, and porous volcanic materials \cite{lahav1978peptide}, \cite{brasier2011pumice}) to wet-dry cycles (hundreds of millions?) involving solutions of varied composition at varied and fluctuating temperatures. In favorable circumstances, these conditions would repeatedly deposit and polymerize films containing amino acids, then return the some of the resulting peptides to aqueous solution.

\subsection{Prebiotic processes produce potential imprinting media}

As is familiar in both kitchens and chemistry laboratories, heating mixtures of organic compounds is apt to yield adherent coatings of insoluble, amorphous polymeric material; indeed, the production of such materials (“tars”) as the main reaction products has been a nuisance in studies of prebiotic chemistries \cite{mccollom2013miller}, \cite{benner2012asphalt}. 

Conditions that induce polymerization of peptides are likely to favor polymerization and cross-linking of imprinting media that are formed in the same chemical milieu and with similar gross chemical functionality (Table \ref{tab:interactions}). Given their observed properties, it would be surprising if prebiotic polymerizable substances did not, on occasion, afford favorable conditions for peptide imprinting (see Section \ref{sec:diverse_media}).

Plausibly-common conditions would also favor surface imprinting. If imprinting media are relatively hydrophobic, amphiphilic molecules (\textit{e.g.,} many peptides)  will tend to segregate to aqueous interfaces, preferentially forming surface imprints \cite{tan2007molecularly}; the preferential formation of surface imprints will also be favored (even ensured) if imprinting molecules embed in films of monomolecular or nanometer-scale thickness \cite{shi2007surface}, \cite{tan2007molecularly}.

\subsection{Racemic precursors are not problematic}
\label{sec:racemic_precursors}

Both proteins and nucleic acids are built of homochiral monomers, but a requirement for pre-existing homochirality under prebiotic conditions would be difficult to satisfy \cite{ruiz2013prebiotic}. Chiral molecules produce chiral imprints in racemic media, however, and those imprints can select and operate on molecules of the same chirality from a racemic solution \cite{morihara1992enzyme}, \cite{kempe1995molecular}. Accordingly, IMT should replicate peptide stereocenters as readily as other structural features, rather than requiring homochirality as a precondition for operation.

Proposed mechanisms that might yield compounds in significant enantiomeric excess (\textit{e.g.,} differential photolysis by circularly polarized starlight \cite{de2011non}, the growth and dissolution of chiral crystals \cite{noorduin2008emergence}, or hypothetical autocatalytic cycles \cite{gleiser2012life}) are of dubious prebiotic realism \cite{avalos2010homochirality}, while prebiotic sources of \textit{diverse, substantially homochiral} monomers are profoundly implausible. The presence of a natural and general mechanism for transmitting chirality in racemic environments thus constitutes a major advantage of IMT proposals.

\section{Toward prebiotic peptide replication: \\Product-directed, imprint-mediated chain extension}
\label{sec:toward_peptide_replication}

Peptide/IMT processes are broadly compatible with what we know of prebiotic environments, and as discussed below, product-directed imprint catalysis seems particularly well-suited to peptide ligation driven by wet-dry cycling. Whether these processes could provide sufficient copying fidelity to support Darwinian selection is a key question, discussed in Section \ref{sec:System-level_considerations}. Table \ref{tab:conditions} summarizes a range of considerations relevant to peptide/IMT models.

\begin{table}[tb]
\begin{center}
\caption{Aspects of anticipated or potential prebiotic conditions, processes, and evolutionary capacity.}
\medskip
\medskip
\includegraphics[width=0.95\linewidth, trim=3cm 1cm 3cm 1cm, clip]{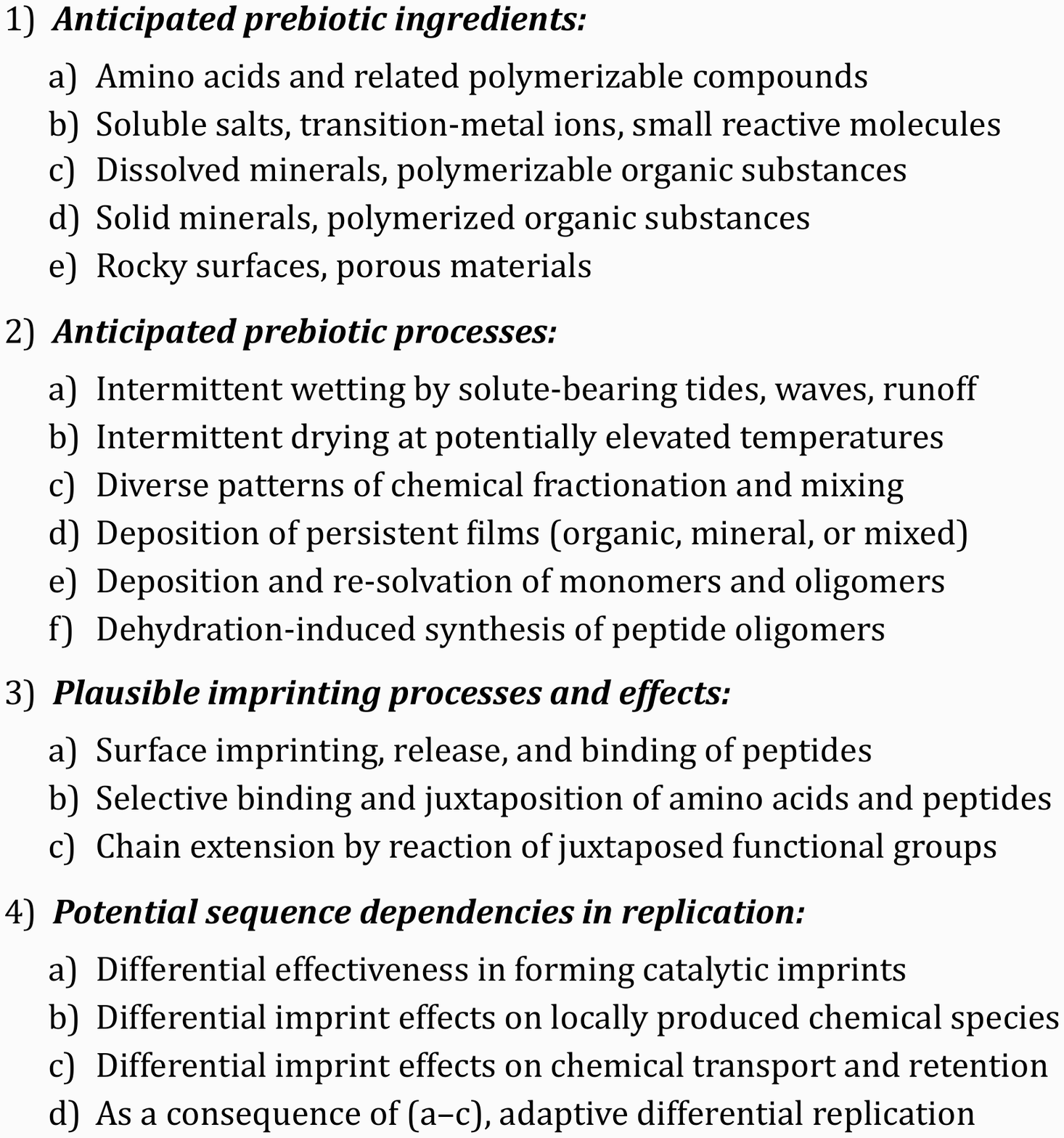}
\label{tab:conditions}
\end{center}
\end{table}

\subsection{Imprints can distinguish between structurally similar peptides}

Experiments have shown that imprint binding interactions can distinguish between peptides that differ by a single amino acid. In one study, a range of single residue substitutions “dramatically lowered binding” \cite{nishino2006selective}; in another, single residue substitutions between alternative hydrophobic side chains lowered $K_\text{a}$ by factors of 13 to 40 \cite{awino2017sequence}. Both imprint formation and peptide binding are typically performed in aqueous media  \cite{janiak2007molecular}\cite{awino2017sequence}, \cite{fa2017peptide}, \cite{fa2018water}, and hence are broadly compatible with prebiotic conditions.

\subsection{Product-directed ligation can likely operate on peptides}
\label{sec:peptide_ligation}

Elementary principles of catalysis suggest that imprint-mediated orientation and juxtaposition of peptides can induce selective bond formation under dehydrothermal conditions by (greatly) increasing the frequency of potentially reactive encounters between complementary amine and carboxylic acid groups. This mechanism is entropic, not enthalpic: Reaction acceleration does not require transition state stabilization, but does require that reactant configurations be compatible in both position and orientation.

Juxtaposition of functional groups by antibody binding can increase their effective molarity to $\gg$\SI{10}{\molar}  \cite{hilvert2000critical}, more than $10^4$ times that of a moderately-concentrated (\SI{1}{\milli\molar}) solution, yielding proportional increases in reaction rates. Antibody catalysis of peptide bond formation and ligation (albeit templated on transition-state analogs) can increase reaction rates by similar ratios \cite{jacobsen1994antibody}, \cite{smithrud1997investigations}.

Binding of reactants to imprints templated on target reaction products would provide an enthalpic driving force for the reaction (in addition to the dominant entropic driving force), but could potentially impair compatibility between the geometries of bound reactants and reaction transition states, increasing transition state energies relative to those of unbound molecules. Inspection of transition-state geometries in the formation (equivalently, hydrolysis) of peptide bonds suggests that geometric compatibility may not be an unduly stringent constraint on reactions between unactivated precursors \cite{xiong2006theoretical}; the flexibility of  polymeric imprinting media can further relax this potential adverse constraints on reaction geometry. Computational modeling could shed light on these questions.

Perhaps surprisingly, simple amine-carboxylic acid condensation reactions driven by wet-dry cycling may be more favorable for imprint-mediated catalysis than reactions between activated precursors in steady-state, solvent-immersed systems: The steric bulk of chemically activated species (\textit{e.g.,} esters, thioesters, and anhydrides, but perhaps not acid chlorides) could potentially interfere with their binding and reactivity in product-templated imprints.

\subsection{Wet-dry cycling can relax trade-offs between binding and turnover}
\label{sec:wet-dry_binding}

Imprints that would fail as catalysts under steady-state conditions could act as effective catalysts in wet-dry cycling. In a steady-state environment, reactants must bind and products be released under identical conditions. Because strong binding to reactants increases reaction barriers, while strong binding to products inhibits product release, effective catalysis under steady-state conditions requires a relatively delicate balance of binding energies that may be unlikely in a system not tuned by design or evolution.

Catalysis driven by wet-dry cycling would be less subject to binding \textit{vs.} turnover trade-offs. Drying drives binding, first concentrating solutes, then eliminating solvent: The question is not \textit{whether} a molecule will bind to an imprinted surface, but \textit{where}. This forced-binding effect can help to compensate for the imperfect complementarity between reactant pairs and product-templated imprints. Conversely, wetting can facilitate the release of products, which need not be tightly bound.\footnote{Binding selectivity across a set of molecules is substantially independent from the general magnitude of binding strengths.} Section \ref{sec:computational_modeling} notes the importance of distinguishing kinetic from thermodynamic control of differential binding in connection with product inhibition under cyclic reaction conditions.

\section{Toward open-ended genetic evolution: \\Faithful copying, selection and the Darwinian threshold}
\label{sec:fidelity} 

For the concepts of “genotype” and “Darwinian selection” to be applicable to a system requires sufficiently faithful replication of information; frequent disruptive substitutions in copying would overcome selective pressures, destabilize genotypes, and preclude Darwinian accumulation of functional structure. Although several considerations suggest that copying fidelity could potentially be high, the anticipated complexity of interaction networks in IMT processes will present challenges in modeling threshold requirements for Darwinian selection. This section first examines considerations at the level of molecular interactions and processes, then turns briefly to system-level questions of replication and selection. Potential modeling strategies will be outlined in Section \ref{sec:computational_modeling}

\subsection{Monomers, environments, and pathways affect copying fidelity}
\label{Monomers_environments_pathways}

Rates of monomer substitution in potential copying pathways will depend not only on the relative binding affinities of sequences of reaction precursors, but also on differences among their concentrations; for example, high concentrations of physically distinctive monomers would tend to support replication with lower substitution rates. Sequential reaction and binding steps afford further mechanisms that could reduce the incorporation of incorrect monomers into longer product sequences.

Note that not all copying error are equivalent: For example, an amino-acid residue \textit{A} might mistakenly template residue \textit{B,} and \textit{B} might mistakenly template \textit{A,} yet the result would be an equilibrium between \textit{A-} and \textit{B-}containing oligomers, rather than open-ended drift in sequence-space \cite{eigen1988molecular}, \cite{iwasaki1999simple}.

\subsubsection*{Structural and chemical differences can favor copying fidelity}
\label{sec:discriminability}

In product-directed ligation mediated by binding  juxtaposition, and distinct dehydrothermal reaction episodes, \textit{binding} specificity should strongly correlate with \textit{catalytic} specificity. As noted above, imprints can strongly discriminate between peptides that differ by a single amino-acid side chain \cite{nishino2006selective}, \cite{awino2017sequence}; the energy differences can exceed those that distinguish correct from incorrect pairings of  bases in an RNA duplex \cite{kierzek1999thermodynamics}.

Greater structural and chemical differences among monomers (and hence downstream oligomers) will increase differences in binding affinities and hence favor copying fidelity. Among biological amino acids, side chains differ in bulk, hydrophobicity, charge, and hydrogen bonding capacity, properties which can induce specific protein folding and function. Amino acids found in model and actual prebiotic environments have a narrower range of side-chain structures, yet show a greater range of backbone structures and patterns of side-chain attachment. The latter differences should potently differentiate their imprint-binding affinities.

Modern genetically-encoded amino acids are homochiral, yet in peptide IMT, mixed chiralities promise not difficulties, but improved copying fidelity. Different sequences of {\small L} and {\small D} residues will tend to be strongly discriminable as a consequence of opposing side-chain orientations, even when those side-chains are identical. For similar reasons, typical imino acids, though achiral, will tend to be strongly discriminable from amino acids as a consequence of their shifted side-chain positions, and different sequences of \textalpha- and \textbeta-amino acids will tend to be strongly discriminable not only as a consequence of their differences in local backbone geometry, but through their effects on the orientations and axial displacements of adjacent sequences. Looking beyond amino and imino acids, diamines and dicarboxylic acids can serve as monomers embedded in peptide chains, yet reverse downstream peptide-bond orientations. All of these strongly-discriminable relatives of familiar {\small L}-\textalpha-amino acids occur in prebiotic reaction mixtures (Section \ref{sec:prebiotic_amino}), and hence are candidate monomers in peptide/IMT processes. Differences in binding affinities across these structural classes should exceed those of base-pair matches and mismatches in nucleic acids.

\subsubsection*{Chemical separation can favor copying fidelity}
\label{sec:chemical_environments}

The fidelity with which an imprint can template the synthesis of a copy of the imprinting molecule will depend not only on properties of the molecule and its imprint, but also on the concentrations of potential precursors in the pool of reactants.

The effectiveness of imprints in affinity chromatography \cite{ekberg1989molecular}, \cite{mosbach1994molecular} suggests mechanisms by which prebiotic systems might mimic purification steps in organic synthesis: Imprints can affect the chemical composition of a local environment by differential retention of molecules across cycles of flushing with solutions derived from potentially diverse upstream sources. Even weaker separation mechanisms (“beach chromatography” \cite{bywater2005did}) will ensure that compositions of raw prebiotic reaction mixtures sometimes differ greatly from the compositions of reactant pools in IMT processes. Processes in which subpopulations of molecular species affect their own retention under wet-dry cycling could potentially amplify the diversity of distinct prebiotic chemical environments (see \cite{guttenberg2017selection}).

\subsubsection*{Sequential process steps can favor copying fidelity}
\label{sec:sequential_steps}

Organic synthesis protocols frequently include purification of reaction products before their use as next-stage reactants in order to prevent by-products from one stage from degrading yields by acting as flawed building blocks in subsequent stages. Separation mechanisms of the sort just discussed could produce a similar effect through fluid flow and differential retention between reaction steps.

A further effect might arise in prebiotic IMT: In sequential processes, misincorporated monomers will tend reduce the binding affinity of their containing oligomers to imprints during subsequent steps, tending to reduce cumulative errors. Thus, imprint-mediated processes can exhibit not only selective inclusion of monomers in oligomers, but also selective exclusion of flawed oligomers from subsequent reactions. The underlying thermodynamic principle---increasing fidelity by coupling a free-energy source to the differential exclusion of mismatched, weakly-bound chemical units---parallels that of kinetic proofreading \cite{hopfield1974kinetic}.

\subsection{System-level considerations determine evolutionary capacity}
\label{sec:System-level_considerations}

Evolutionary capacity is a soft concept that entails replication of populations of sequences that are shaped by selection pressures and focused in sequence space. Although evolutionary capacity requires that selection pressures counter the tendency of monomer substitutions to cause sequence diffusion, it need not require focusing to a single sequence \cite{kauffman2000investigations}). Selection pressures and system-level fidelity are thus crucial, but trade off against one another. System-level fidelity in an IMT process will depend on sequences and networks of templating reactions and ligations, processes which cannot be modeled as an error rate per monomer in a sequential polymerization process. In the end, for evolutionary capacity to be of interest in abiogenesis, IMT processes must support the elaboration of complexity in directions that support the emergence of structural and functional complexity, the subject of Section  \ref{sec:metabolic_evolution}.

\subsubsection*{Selection pressures act on evolvable peptide/IMT systems}
\label{sec:selection_pressures}

Selection pressures inevitably act through the \textit{direct phenotype} of a sequence, the interaction between a pre-existing environment and physical features that directly correspond to the sequence itself. For example, natural selection would act from the start on both the composition and sequence of templating oligomers: In a milieu that supports product-directed imprint catalysis, selective pressures will tend to favor oligomers that are effective in forming templates that are both \textit{productive} and \textit{faithful} in that same milieu.

Even without metabolic complexity, environmental compartmentation (Section \ref{sec:localization}) can induce selection that operates through the \textit{indirect phenotype} of a sequence, mediated by effects of the sequence on its own surroundings. It what is perhaps the simplest example, the imprints would tend to retain (partially-)matching sequences though wet cycles, while the retained sequences would tend to produce (partially-)matching imprints. This retention-mediated feedback mechanism would favor focusing in sequence space (around one or more sequences), providing a general pressure toward sequence homogeneity---in effect, higher replication fidelity---while allowing sequences to evolve in response to functionally relevant selective pressures.

\subsubsection*{IMT-mediated replication pathways are potentially complex}
\label{sec:complex_replication}

Conventional polymerization through single-monomer addition is within the potential repertoire of IMT processes, yet other mechanisms seem likely to operate (even dominate) in realistic networks of reactions and imprint formation.

In particular, the discussion above has referred to oligomer ligation as a mechanism for replicating sequences, yet products (even with perfect local copying) need not be the same \textit{length} as the original. An imprint could readily direct the ligation of oligomer-pairs that incorporate contiguous subsequences of the imprinting oligomer, yet differ overall: Bound reactants could be shorter (only partially filling the imprint), or longer (hence only partially bound), and so on. An oligomer could thus direct the formation not only of copies, but of a range of products that contain subsequences of the parent.

\subsubsection*{The requirements for copying fidelity invite investigation}
\label{sec:systemic_complexity}

As a consequence of novel mechanisms, results from studies of threshold fidelity requirements in nucleic acid systems are not directly applicable to IMT. In the midst of unavoidably more complex processes, with many potential patterns of chain extension, ligation, mutation, recombination, retention, \textit{etc.,} the  threshold requirements for effective genetic processes are as yet unclear. The identification of conditions that support the elaboration of genetic complexity in systems with these characteristics would best be pursued through computational simulations (Section \ref{sec:system_modeling}).

\section{Toward open-ended metabolic evolution: \\ Macromolecules, localization, and compartments}
\label{sec:metabolic_evolution}

The preceding section completes an outline of the basic peptide/IMT model, a potential mechanism for Darwinian evolution of functional molecular complexity in realistic prebiotic environments. This basic model falls short of describing what can be considered an actual biology, lacking structures and processes comparable to those of living cells. To be credible, a model for abiogenesis should at least \textit{suggest} paths in this direction, toward high-fidelity genetic processes, complex metabolism, and complex functional structures.

The present section explores processes and macromolecular architectures (\textit{e.g.,} branch\-ed polymers) of kinds that might naturally emerge from an effective but rudimentary IMT-based evolutionary process, considers their relevance to molecular functionality, then explores the potential emergence of compartmented and even cellular systems that embody more extensive structural and metabolic complexity. A key theme will be the evolutionary coupling of genotypes (structural information \textit{per se}) to phenotypes that are not fully embodied in the physical properties of the genetic molecules themselves. A key principle of inquiry will be to minimize the invocation of mechanisms would require more than incremental elaboration of peptide/IMT processes and products.

It is important to recognize the role of this discussion: Proposed mechanisms for peptide/IMT-based genetic processes stand or fall independent of \textit{specific} proposals for paths toward complex systems. Proposals for paths forward do not add complexity to the basic proposal, but instead suggest potential connections between unfamiliar surface-bound, imprint-mediated chemistries operating in amorphous media and more familiar biological architectures in which replication, metabolism, and evolutionary competition operate on structured molecular entities in solution-phase systems. The goal is to show that there are no fundamental discontinuities between simple peptide/IMT systems and functionally-advanced systems of the kind that could in principle support the emergence of nucleic-acid based genetics.

\subsection{IMT does not privilege linear, monomer-by-monomer synthesis}
\label{sec:nonlinear}

The argument for accessible, incremental elaboration of metabolism begins by examining potential directions for elaboration of functionality at the level of molecular structure.

\subsubsection*{Branched and cyclic structures can be directly replicated}

Imprints are not polymerases, and imprint-mediated templating should favor neither monomer-by-monomer synthesis nor linear products. Indeed, because trifunctional peptide monomers with branching capacity (\textit{e.g.,} aspartic acid, glutamic acid, and \textalpha,\textgamma-diaminobutyric acid) are present in prebiotic reaction mixtures \cite{ring1972prebiotic}, branched products are apt to be common. Branching can improve functional capacity because branched topologies constrain the space of accessible molecular conformations, reducing entropy and increasing the frequency of intramolecular interactions; as noted below, branched structures have a greater propensity to exhibit protein-like behaviors. Imprint-directed cyclization \cite{cenci2016guided}, \cite{shen2016catalytic} could further reduce the entropy of templated peptide products.

\subsubsection*{Convergent assembly could shorten reaction sequences}

Organic chemistry often employs convergent synthesis, and in IMT driven by discrete wet-dry cycles, convergent reaction pathways would be natural and perhaps difficult to avoid. An idealized convergent synthesis of a peptide oligomer would begin with amino acid dimerization and then proceed through a series of ligations of fragments of similar length. In this model, an oligomer of length $N$ is assembled by ~$N$ ligations that require only ~$\log_2 N $ reaction cycles. Regarding replication fidelity, a convergent pathway could potentially expose each misincorporated monomer to a series of ~$\log_2 N$ opportunities to be “recognized” as such and, under suitable conditions of wet-phase molecular transport, excluded from subsequent reactions (see Section \ref{Monomers_environments_pathways})

\subsubsection*{Ligation by short-sequence imprints can link longer sequences}

Imprints can bind short sequences of longer peptides, a principle employed in applications of imprinting to protein separation \cite{nishino2006selective}. As noted in Section \ref{sec:System-level_considerations},  imprints could potentially bind and ligate polymers not by matching and binding their entire lengths, but by binding only segments adjacent to the point of ligation. This process resembles homologous recombination, but also provides a mechanism for generating products that are far longer than the imprinting oligomers, while embedding their sequence information. The ability to recombine and extend structures in this fashion would facilitate exploration of the space of potential macromolecular functionality.

\subsection{Localization of effects is crucial to evolving metabolic complexity}
\label{sec:localization}

In single-celled organisms, a gene’s effects are coupled the gene’s replication through the confinement of gene products and metabolites within a cell membrane. In IMT processes, functionally analogous localization follows naturally from the general properties of polymers, surfaces, and the interaction of wet-dry cycles with surface topography, enabling the elaboration of metabolic complexity without requiring the prior appearance of membranes.

\subsubsection*{Localization of effects links genotypes to indirect phenotypes}

In cells, genetic molecules can affect their cytoplasmic environments through multiple routes, and the resulting indirect (\textit{e.g.,} metabolic) phenotypes then can influence the replication of the genes themselves.
 
As illustrated by viruses, the evolutionary coupling of a genotype to its indirect phenotype need not depend on cell division; rather, the key characteristic is the ability of a genetic molecule to differentially affect \textit{its own} replicative success through its influence on an immediately surrounding environment. By contrast, if the effects of genetic molecules were solely embodied in molecules free to escape by diffusion, genotypes would not be subject to selective pressures mediated by these molecules, and accordingly, there would be no selective pressure toward the evolution of corresponding metabolism \cite{cantine2018environmental}.

From the perspective of modern biologies, it is natural to identify molecular localization with cell membranes. Surface-bound peptide/IMT models, however, not only suggest membrane-free mechanisms for molecular localization, but can exploit wet-dry cycling as a free-energy source, avoiding difficulties that arise in models that postulate membrane compartmentation as a basis for initial energy metabolism \cite{ruiz2013prebiotic}.

\subsubsection*{Fixed surfaces localize both imprints and anchored molecules}

Unlike soluble chemical species, imprints do not diffuse, and soluble polymers can be anchored to surfaces by any of a range of mechanisms, including noncovalent binding to imprints and covalent binding to the surface itself. The inherent immobility of imprints does not impair their function, while anchoring a polymer at a point or segment leaves other portions free to interact with surfaces, imprints, and molecules within a bounded radius.

Catalytically active imprints of mobile segments of an anchored polymer would be clustered and could contribute to the synthesis of additional (potentially locally anchored) polymers that share sequence information with the first. In addition, such imprints could contribute to the extension of anchored polymers by appending or adding as branches shorter sequences. Thus, one can envision (or perhaps expect) the production of bushy, surface-anchored macromolecular polymers in which multiple copies of diverse, replicable substructures produce joint effects through interactions mediated by local imprints, molecular neighbors, or intramolecular encounters.

In this picture, surface anchoring plays a role that is in some respects comparable to membrane compartmentation, while fragmentation and re-anchoring of polymers could play a role comparable to cell division or viral replication. None of these phenomena (aside from chemically-trivial surface anchoring and chain scission) requires mechanisms beyond those inherent in an effective IMT process. Note that this model does not call for the novel chemical species, membrane assembly processes, membrane-content packaging, and solution-phase energy sources required in typical membrane-based compartmentation scenarios. The surface-anchoring model does not, however, provide for compartmentation of diffusible metabolites.

\subsubsection*{Surface topography can confine diffusible metabolites}

Although surface anchoring can effect localization of polymers, membranes in modern biology play a crucial role in confining diffusible, small-molecule metabolites. This further localization of genetic effects appears essential to the evolution of small-molecule metabolism, but surface topography can provide confinement mechanisms without membranes.

\begin{figure}[tb]
\begin{center}
\includegraphics[width=0.8\linewidth, trim=1cm 5cm 1cm 5cm, clip]{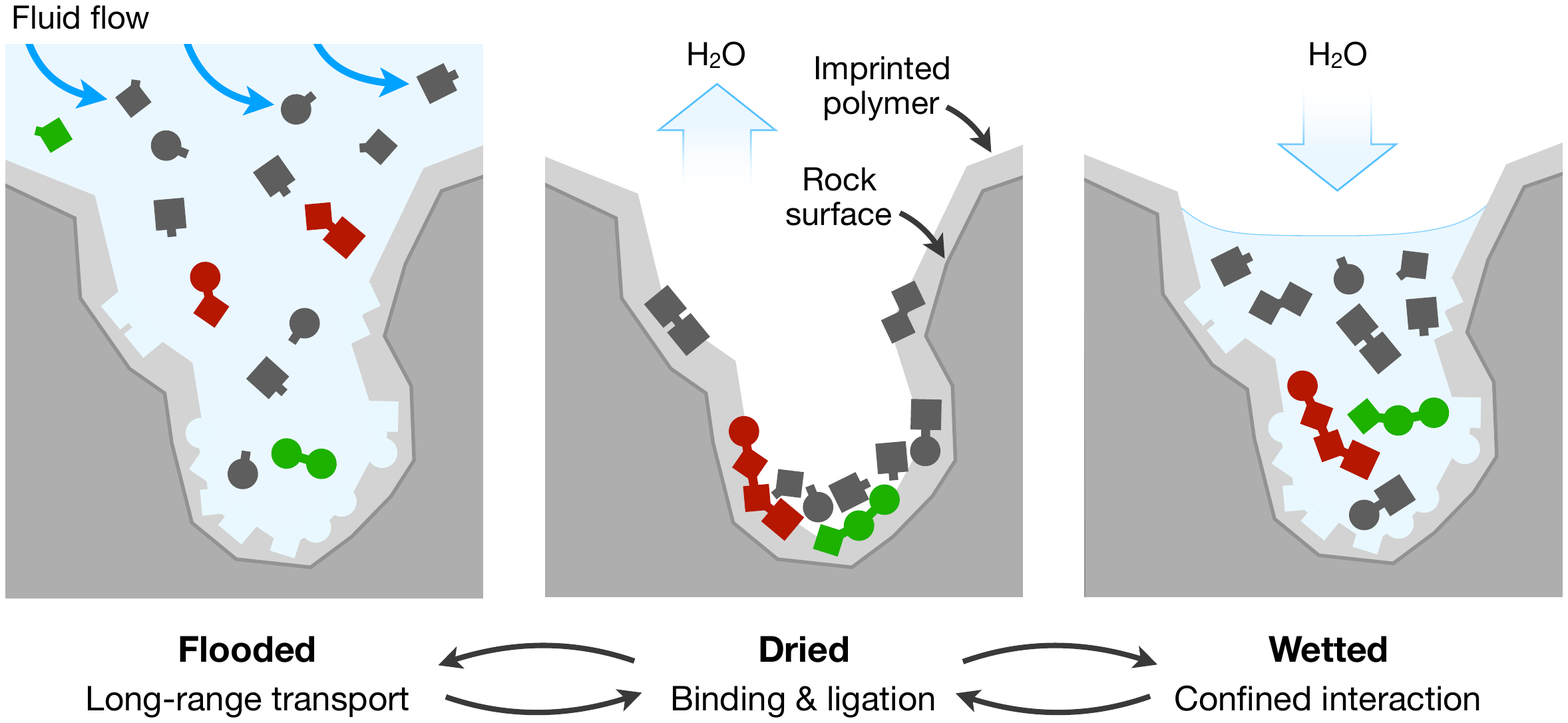}
\caption{Schematic diagram of fluid compartmentation and exchange through intermittent flooding of pore-bearing surfaces. Surface pores can be dry, separately wetted, or in communication with a continuous aqueous phase (“flooded”). In the dry state, dehydration drives binding and ligation of oligomers (Section \ref{sec:wet-dry_binding}). In the flooded state, pores exchange soluble contents with an extended region, while In the wet state, soluble molecules are mobile, yet confined to the pore volume. In conjunction with selective surface binding and retention of small molecules, alternation of wet and flooded states could substantially mimic the functionality of a semi-permeable membrane.}
\label{fig:wet_and_dry}
\end{center}
\end{figure}

On rough or pore-bearing surfaces, wet-dry cycles would result in intermittent formation of confined aqueous compartments (see Figure \ref{fig:wet_and_dry}). Like a semi-permeable membrane, wet-dry cycles operating on pore-bearing surfaces can effect both confinement and exchange of diffusible molecular species; a further range of biological membrane functions (\textit{e.g.,} selective permeability, to say nothing of active transport and chemiosmotic coupling \cite{lane2010did}) would of course be absent. In this scenario, replication and dispersion of the genetic material would involve, not the division and proliferation of membrane compartments, but the transfer of competing replicable molecules between pre-existing pores, a process analogous to the transfer of viruses or plasmids between cells.

\subsubsection*{Metabolic specialization suffices to explain the puzzle of homochirality}

As discussed in Section \ref{sec:racemic_precursors}, the peptide/IMT model provides a mechanism for the emergence of replication and metabolism in racemic environments, obviating the challenging yet widely assumed requirement for prebiotic homochirality. Beyond this, however, peptide/IMT (in conjunction with confinement of diffusible metabolites) could readily give rise to metabolic pathways specialized around monomers with shared chirality (\textit{e.g.,} {\small L}-amino acids), providing a natural explanation for the emergence of biological homochirality through spontaneous symmetry breaking in metabolic evolution \cite{ribo2017spontaneous}.\footnote{Metabolism, potentially in the absence of cells, could explain the otherwise puzzling, consistent reports of {\small L}-amino acids found in enantiomeric excess in carbonaceous chondrite meteorites (up to 2-fold {\small L}/{\small D} enrichment \cite{sephton2002organic}, \cite{aerts2016contamination}); these objects are derived from formerly-wet, potentially life-supporting \cite{abramov2011abodes} parent bodies that predate Earth’s late-Hadean era. The absence of a range of conventional biomarkers (“molecular fossils”) among the products of a simple metabolism would be unsurprising.}

\subsection{IMT suggests incremental paths to complex metabolism}

Incremental elaboration of imprinting mechanisms can lead to enzyme-like functionality and complex metabolism.

\subsubsection*{Templated peptides could enrich the functionality of imprinting media}

Embedded peptide sequences can contribute structure to what are nominally amorphous imprinting media, potentially inducing functionally-significant modifications of imprint geometry, binding specificity, and catalytic activity. By forming part of an imprint, positioned and oriented through interactions with the imprinting molecule, a templated peptide product could augment imprint functionality without itself providing sufficient binding affinity to act as an effective template, receptor, or catalyst---functions that in modern biology typically require solution-phase folding of relatively large molecules. Imprint-embedded peptide structures selected for specific chemical functionality would have the capacity to introduce a range of enzyme-like catalytic mechanisms. In a limiting case, local imprinting media would themselves consist entirely of aggregates of templated products.

Analogous systems are described in the imprinting literature: Researchers have structured imprinting media by incorporating oligopeptides \cite{yoshikawa1998molecularly}, macrocyclic hosts \cite{dickert1999molecular}, \cite{lay2016state}, and polydentate chelators \cite{singh2000towards}, \cite{becker2004exploiting}, \cite{tada2010molecular}; catalysis has been enhanced by incorporating transition-metal complexes \cite{severin2000imprinted}.

In the scenario outlined above, functional peptide sequences are replicable by imprinting, but typically augment functionality when embedded in imprints that operate on \textit{other} sequences. For a sequence to facilitate the replication of a different sequence would constitute an indirect phenotype, hence its evolutionary selection requires physical localization along the lines discussed above.

\subsubsection*{Peptide-IMT evolution opens paths to regulated enzyme-like functionality}

Peptide/IMT genetic mechanisms that cross the Darwinian threshold could support the elaboration of functional polypeptides and related macromolecules without the prior emergence of a ribosomal translation system, providing an incremental evolutionary path to macromolecular structures that provide enzyme-like functionality. Further, as noted above, even small peptide oligomers could provide elements of protein-like functionality when embedded in imprints that provide requisite binding strength and specificity. From an evolutionary perspective, there is no sharp dividing line between imprint-embedded peptides and independently folded, protein-like structures.

Although branched peptides differ from proteins, branching tends to facilitate---not impair---independent folding and protein-like behavior. Even when disordered, hyperbranched structures can exhibit protein-like activity \cite{mamajanov2017protoenzymes}. Further, by constraining available conformation space, branched and cyclic topologies reduce entropic barriers to ordered folding \cite{mutter1992template}. Even without exploiting non-linear topologies or non-biological structural diversity, protein designers have constructed foldable sequences based on prebiotic amino acids \cite{longo2013simplified}.

The functionality of independently folded peptide/IMT products would be determined by their evolved structures, and such products could play the role of enzymes in every important respect. There is then there no restriction to product-directed catalysis, and evolutionary processes could explore a full range of enzymatic catalytic mechanisms. Independently folded structures would constitute a pre-adaptation for solution-phase metabolism.

Metabolism in modern biology is regulated by numerous signaling and feedback pathways, and a genetic system with the capacity to produce independently folded structures could exploit a similar range of regulatory mechanisms.

Note that independently-folded structures need not be free to diffuse. An extensive literature describes the immobilization of enzymes and other macromolecules by linking to surfaces with preservation of function \cite{klibanov1983immobilized}.

\subsubsection*{Wet-dry cycles could activate chemical species for solution-phase metabolism}

In the basic peptide/IMT model, externally imposed wet-dry cycles act directly on mono\-mers and oligomers to drive polymerization. Modern biology, by contrast, employs activated chemical species to drive similar processes---and many others---at the pace of enzyme reaction cycles. Continuing the theme of expanding  the model toward biological functionality while minimizing additional mechanism, it is of interest to consider how wet-dry cycles might provide reservoirs of activated chemical species, and do so by a volumetrically distributed process.

Wet-dry cycles can serve as potent sources of free energy in molecular systems: Concentrating a solution by a factor of $10^4$ (equivalent to evaporating a \SI{10}{\micro\metre} film to a thickness of 1 nm) shifts $\Delta G$ toward binding by ~\SI{20}{\kilo\joule} per mole of solute, while in lysozyme crystals, reducing ambient humidity from 90\% to 60\% increases the mechanical-deformation component of the induced $\Delta G$ by ~\SI{65}{\kilo\joule\per\mole}, \cite{morozov1988interpretation}. This value is comparable in magnitude to the free energies of relevant biomolecular bond formation and hydrolysis reactions; for example, in hydrolysis of typical thioester, ester, and peptide bonds, and of ATP, the values of –$\Delta G^{\circ\prime}$ are respectively $\mathtt{\sim}$30, $\mathtt{\sim}$20, $\mathtt{\sim}$10, and $\mathtt{\sim}$30 \SI{}{\kilo\joule\per\mole}. ATP synthase couples proton flow to ATP synthesis through mechanical deformation of active sites in the F1 complex \cite{wang1998energy}; dehydration-induced mechanical deformation of protein-like structures could presumably be harnessed to the production of chemically activated species in a similar manner.

Thus, one can readily envision enzymatic functionality that binds and activates small molecules through drying, then releases the products to enable multiple steps of solution-phase chemistry during a single wet phase. In particular, enzymatically assisted dehydrothermal synthesis could produce amino-acid esters, a class of molecules that is not only widely used in laboratory peptide synthesis, but also the basis for ribosomal peptide synthesis (acylated tRNA molecules are amino-acid esters). Dehydrothermal esterification in this model could potentially occur at relatively high volumetric rates (>$10^7$\SI{}{\per\cubic\micro\metre} per cycle) in protein-like structures distributed throughout a thick film of material, providing a substantial supply of activated chemical species to a co-confined solution-phase metabolic system.

Parallels with modern peptide synthesis can be taken further: Peptide chemists commonly employ methods in which peptide chains are retained by anchoring to solid-phase structures though cycles of amino-acid addition and solution interchange, while ribosomes add amino acids and transfer peptide chains between what are, at least transiently, anchored tRNA molecules. An analogous process based on prebiotically-available chemical species would both activate and bind amino acids through dehydrothermal esterification of hydroxyl groups (\textit{e.g.,} of serine residues) in locally retained peptide structures, followed by transfer of the activated amino acids to peptide chains that are likewise locally retained. Such a process would enable localization of what is in effect small-molecule metabolism without requiring full compartmentation. Through a different choreography of acylation reactions, processes based on similar ester/amine chemistry could mimic the swinging-arm mechanism of nonribosomal peptide synthetases \cite{mootz2002ways}. 

\subsubsection*{Membrane compartmentation could develop incrementally}

Although perhaps requiring additional molecular mechanism (beyond peptides and imprinting media), one can envision membrane-dependent functionality developing initially in pores capped by a semipermeable material (proposed in connection with molecular imprinting by Lauterbur \cite{lauterbur2008spontaneous} ), without requiring the single-step emergence of full membrane-based compartmentation and cell-like proliferation. Even here, however, peptide-centric mechanisms seem plausible: Peptide-based membrane functionality has been proposed in the context of abiogenesis \cite{childers2009peptide}, \cite{egel2009peptide}, while in the laboratory, assemblies of broadly peptide-like hyperbranched polymers have shown membrane-like properties \cite{zhou2004supramolecular}, \cite{jiang2015hyperbranched}. Given an effective base of genetics and metabolism, one can readily envision incremental evolution from semipermeable capping materials to semipermeable membranes through substitution of molecular components.

Accordingly, it seems likely that an effective surface-dependent metabolism could support the incremental emergence of membrane-bound systems, while membrane-based small-molecular confinement could broaden the scope of metabolic processes. The ability of some modern organisms (even eukaryotes) to survive desiccation indicates that continued exploitation of wet-dry cycling and dehydrothermal reactions in metabolism is compatible with the emergence of full membrane compartmentation.

\subsection{Peptide systems could support surface-independent replication}
\label{Solution-phase_genetics}

It is interesting to consider whether surface-based IMT genetic mechanisms could give rise to surface-free (“cytoplasmic”) replication without first elaborating a PMT process. A negative answer would not undercut the basic concept of peptide/IMT abiogenesis, but a positive answer would extend the scope of potential IMT mechanisms into the domain of recognizable cellular life.

\subsubsection*{Coding-capacity constraints present theoretical difficulties}
\label{coding_capacity}

Any independent genetic system must somehow encode the information embodied in its associated machinery. Modern genetic systems can employ machinery of modest functional complexity to provide indefinitely large coding capacity because nucleic acid polymerases operate uniformly on all genetic sequences. Proposed surface-bound IMT processes, by contrast, employ structures that perform \textit{sequence-specific} monomer addition and oligomer ligation, and in effect encode information in the products themselves. For an analogous process to operate without surface imprinting might seem to require the synthesis of a great number of sequence-specific catalysts, collectively embodying sequence complexity that is beyond the coding capacity of the sequences they can copy.

This problem of system complexity \textit{vs.} coding capacity was first articulated in early theoretical biology \cite{hausmann2013grasp}: How could a set of $N$ enzymes, each with a fixed sequence specificity, direct the synthesis of (at least) that same set of enzymes, given that these would collectively comprise sequences containing on the order of $100N$ specific, sequence-defining peptide bonds?

\subsubsection*{Pauling-style imprinting could circumvent coding-capacity constraints}
\label{Pauling-style_genetics}

To address the even knottier problem of antibody diversity, where the number of distinct antibodies greatly exceeds the number of germline genes, Pauling proposed what amounts to a molecular imprinting model in which polypeptide chains (in immunoglobulins) “[Have] accessible a very great many configurations with nearly the same stability; under the influence of an antigen molecule they assume configurations complementary to surface regions of the antigen\ldots” \cite{pauling1940theory}.

The facile production of molecular imprints having antibody-like binding properties shows that Pauling’s idea, although not realized in nature, was sound in its chemical fundamentals. Supplementing this scheme with a relatively promiscuous cross-linking mechanism to consolidate the configuration of an “antibody” would yield a solution-phase process in which the “imprinting medium” comprises a number of distinct, genetically defined polymers that are adapted to \textit{partially} differentiated roles, gaining further, differentiated information through post-synthesis imprinting. Note that solution-phase, protein-scale structures can, in fact, serve as peptide imprinting media \cite{hoshino2008peptide}, \cite{fa2017peptide}.

Translating this idea to an IMT context, a moderate number of labile, semi-structured macromolecules, each produced by a series of imprint-mediated ligations, would though subsequent imprinting serve as an indefinitely large number of sequence-specific ligation catalysts. In rough analogy to ribozymes in proposed RNA-world models \cite{rich1962problems}, \cite{gilbert1986origin}, \cite{bernhardt2012rna}, \cite{neveu2013strong}, genetic molecules in a solution-phase peptide/IMT system would serve both as \textit{templates} for transmitting their sequence information and as \textit{catalytic components} active in copying other sequences.

From an evolutionary perspective, template-directed copying differs qualitatively from code-directed ribosomal translation. Codon reassignments have been phylogenetically rare since the last universal common ancestor \cite{koonin2017origin}, while IMT-based mechanisms would be relatively fluid in their ability to exploit novel monomers: Imprinting a structure containing novel monomers need not require new mechanisms.

\subsection{Emergent complexity is a requirement, not a hindrance}

This section has suggested potential incremental paths from primitive surface-bound genetic processes to complex solution-phase metabolism and genetics, while remaining within the mechanistic scope of peptide/IMT chemistries. In this context, the criteria for plausibility are not those of simplicity \textit{per se,} but of economy of mechanism in explaining increasing complexity. The incremental elaboration of  complex functionality is an essential aspect of any model of the emergence of life; the theoretical question is the extent to which a model could support this as a natural and incremental process.

\section{Toward answering the hard questions: \\Experimentation and computational modeling}
\label{sec:experimental_and_computational}

Both experimental and computational studies can contribute to our understanding of potential peptide/IMT processes, experimental studies primarily by addressing critical uncertainties regarding the potential scope and fidelity of product-directed peptide ligation, and computational studies primarily by exploring conditions under which networks of IMT processes could lead to peptide replication. In both instances, natural research strategies would first seek idealized conditions that yield positive results, then seek conditions that yield positive results while better satisfying the constraints of prebiotic realism.

\subsection{Experimental studies}
\label{sec:experimental}

Peptide/IMT models suggest novel lines of experimental inquiry, first to explore the feasibility of critical mechanisms under optimized laboratory conditions (selected monomers, tailored imprinting media, \textit{etc.}) and then to explore their feasibility and fidelity in more realistic models of prebiotic environments. In both contexts, key proof-of-principle demonstrations would include imprint-mediated, product-directed catalysis of:

\begin{itemize}[itemsep=0pt]
\item Amino-acid dimerization.
\item Chain extension by monomer addition.
\item Chain ligation.
\end{itemize}

Identifying conditions that result in high-fidelity copying against a background of mismatched oligomers would be highly significant. Understanding of the potential generality of peptide/IMT processes would be improved by further demonstrations: 

\begin{itemize}[itemsep=0pt]
\item Ligation of long chains by short-chain imprints.
\item Ligation of short chains by long-chain imprints. 
\item Templating of branched and cyclic peptides.
\item Templating with racemic monomers (perhaps from the start).
\item Templating with monomers other than \textalpha-amino acids.
\end{itemize}

\subsubsection*{Toward prebiotic realism}

Given the lack of success to date in demonstrating plausibly-prebiotic nucleic-acid chem\-istries and PMT, almost any experimental demonstration of copying peptide-sequence information under qualitatively-realistic prebiotic conditions would constitute a major milestone in experimental studies of abiogenesis.

As a general rule, the conditions of greatest fundamental interest will be chemically \textit{plausible} yet functionally \textit{favorable.} Although realism calls for environments and chemistries consistent with credible models of  early-Earth conditions, it is reasonable to postulate both particularly favorable patterns of local environmental cycling and substantial, favorable separation and concentration of chemical species from raw prebiotic mixtures (Section \ref{sec:chemical_environments}). Realistic temperatures are only weakly constrained: Conditions prevailing in late-Hadean environments could presumably range or cycle from moderate (compatible with modern living organisms \cite{Schwartzman2004hyperthermophilic}) to high (driven by volcanic heat sources).

Common peptide synthesis chemistries are of dubious relevance to prebiotic imprint-mediated processes. Activated precursors and uniform reaction conditions are standard in laboratory peptide chain extension and ligation, yet seem unlikely to be successful in modeling imprint-mediated dehydrothermal peptide bond formation. As noted in Section \ref{sec:wet-dry_binding}, wet-dry cycling can relax a range of constraints by facilitating reactant binding and product release, while considerations of transition-state geometry may well favor dehydrothermal reactions employing \textit{unmodified} carboxylic-acid functional groups over reactions employing, for example, active esters (Section \ref{sec:peptide_ligation}). Prebiotic realism of course strongly favors the dehydrothermal approach.

Common salts are realistic components of prebiotic mixtures, and can markedly enhance peptide ligation under dehydrothermal conditions (\cite{rode1999combination}). A range of metal ions can likewise exhibit catalytic activity and are plausible components of seawater and evaporites. Leaching from local mineral deposits or volcanic ash could deliver relatively rare metals to reaction environments \cite{smith1982leachability}.

Potential imprinting materials and substrates span a wide range of compositions and physical forms, including high surface-area granular materials that could diversify local reaction environments while increasing volumetric yields in laboratory apparatus. Organic, mineral, and mixed-composition imprinting media have been employed in experiments, and a range of such materials would be expected in prebiotic environments. Surface imprinting will be favored by thin-film imprinting media, and evaporation of dilute solutions of precursor materials can produce such films. Thermal and perhaps photochemical cross-linking are plausible prebiotic mechanisms for consolidation of imprinting media; evaporites and metal-ion catalysis are again relevant.

In the pursuit of replication fidelity, the most effective experimental methodologies will likely combine model reactants chosen to facilitate sensitive product assays (\textit{e.g,} by alternative ligations forming discriminable FRET pairs) with setups that facilitate parallel exploration of diverse imprinting and reaction conditions. Although reproducibility will be essential, experimental work to date suggests that detailed structural and mechanistic characterization of imprint-based catalysis may be both difficult and unnecessary.

Crossing the Darwinian threshold, the key step in abiogensis, requires surpassing some threshold level of replication fidelity in imprint/ligation sequences. The threshold itself is presently difficult to characterize (Section \ref{sec:computational_modeling}), but a range of metrics can be applied to measure degrees of success and guide experimentation.

\subsection{Computational modeling}
\label{sec:computational_modeling}

Computational modeling could be applied to describe both molecular interactions and potential system-level dynamics. These forms of modeling are, however, substantially distinct.

\subsubsection*{Modeling molecular interactions}

At a molecular level, computational experiments could potentially provide guidance in selecting imprinting media and degrees of consolidation. In particular, increases in the rigidity of a consolidated imprinting medium (equivalently, constraints on induced fit) will tend to increase binding selectivity while restricting the mobility of functional groups during ligation. Molecular dynamics methods may be useful in exploring the resulting trade offs, as well as trade offs in imprint depth, where deeper embedding of a templating molecule may improve specificity at the expense of reducing exchange rates.

\subsubsection*{Exploring potential system-level dynamics}
\label{sec:system_modeling}

The potential genetic capacity of peptide/IMT systems could be explored by studies of system dynamics able to capture the key physical interactions of molecular binding, ligation, release, and transport---together with imprint formation and destruction---under conditions of wet-dry cycling in heterogeneous environments. Simulation of this range of processes calls for stochastic modeling at a moderately high level of physical abstraction \cite{walker2012universal}.

A range of computational techniques, developed under the rubric of “agent-based modeling and simulation” (ABMS), has been applied to analogous problems in molecular systems biology \cite{soheilypour2018agent}. ABMS treats systems as collections of “agents” defined by (\textit{e.g.}) their types, locations, mobility, state changes, and interactions. ABMS techniques are well suited to capturing phenomena with many kinds of entities (here, many potential peptide sequences and their corresponding imprints), spatial localization and transport, and statistical behavior resulting from interactions among small numbers of individuals. The latter could play a critical role in IMT system dynamics: For example, a strongly-binding molecular species cannot block a binding site if no such molecule happens to be present in a given pore during a given wet-dry cycle.

To model this range of range of phenomena is beyond reach of classical methods that integrate differential equations describing concentrations and reaction kinetics. Studies in this tradition have offered insights into molecular-level evolutionary processes (see \cite{eigen1988molecular}, \cite{iwasaki1999simple}), but cannot model processes that involve indefinitely large numbers of chemical species and small numbers of individuals---processes well within the scope of ABMS techniques.

In considering conditions that might have given rise to genetic systems in prebiotic environments, we can think of nature as having conducted a blind search across a high-dimensional parameter space that spans varied surfaces, temperatures, and chemical-mixture compositions, as well as varied temporal patterns of temperatures and chemical exposures. The resources available in nature’s search were planetary or even cosmic in scale. To find conditions for peptide/IMT genetics within reasonable computational resources may require heuristic search methods (\textit{e.g,} evolution strategies \cite{beyer2002evolution}) that are initially applied to idealized models that abstract strongly from realistic prebiotic conditions. At present, we cannot claim to have a working understanding of the patterns of imprint formation, fragment assembly, and chemical transport that would emerge in peptide/IMT processes, much less the requirements for IMT-based sequence replication. The low-hanging fruit for inquiry today is likely to be found through models that capture qualitatively realistic patterns of structure and process without pretense of modeling specific molecular species.

A successful research program of this kind would lead to a convergence between mechanisms demonstrated in chemical experiments and system behaviors that yield genetic systems in simulation. Further along this path might be experimental demonstrations of the emergence of genetic systems in credible models of prebiotic systems.

\section{The scientific context: \\Potential answers to long-standing questions}
\label{sec:long-standing_questions}

Broadly speaking, studies of abiogenesis can either look back from the intricate machinery of present biology, seeking simpler precursors, or can look forward from the tars and chemical broths of prebiotic chemistry, seeking paths to genetic systems and metabolic complexity. Studies following the first approach have centered on the “strong RNA world hypothesis” \cite{rich1962problems}, \cite{gilbert1986origin}, \cite{bernhardt2012rna}, \cite{neveu2013strong}, which proposes that RNA played both catalytic and genetic roles in a prebiotic environment; the challenge has been to explain how such a complex chemistry could have arisen under prebiotic conditions. Studies following the second approach have established that a range of prebiotic processes give rise to biologically relevant molecules, including amino acids and peptide oligomers; the challenge has been to explain how peptide chemistries could have given rise to genetic mechanisms.

Although lines of research inspired by the RNA-world hypothesis have sought prebiotic processes that could produce nucleic acids, no realistic prebiotic model has produced a measurable trace of any nucleic acid monomer, to say nothing of a nucleic acid polymer \cite{orgel2004prebiotic}, \cite{benner2012asphalt}, \cite{bernhardt2012rna}  \cite{mccollom2013miller}, \cite{ruiz2013prebiotic}, \cite{luisi2015chemistry}. RNA-first models nonetheless assume (or tacitly require) the prebiotic existence of specific combinations of substantially-pure nucleic acid monomers at high concentrations, conditions that would still fail to produce nucleic acid polymers \cite{luisi2015chemistry}. Benner has described RNA as “a prebiotic chemist's nightmare” \cite{benner2012asphalt}.

Proposed RNA analogs that replace ribose with simpler backbone components (reviewed in \cite{hud2013origin}) mitigate few of these difficulties. It seems that any viable PMT model (whether based on nucleic acids or not) would require the occurrence of a delicate balance of physical and chemical properties in a specific, complex, self-complementary polymer molecule that is somehow abundant in a prebiotic environment.

Sutherland \cite{sutherland2016origin} summarizes and cites a set of widely argued yet conflicting propositions as follows:

\begin{enumerate}

\item “Darwinian evolution needs informational molecules, so RNA must have come first.” \cite{joyce1999prospects}

\item “You can’t get by without building blocks and energy, so metabolism must have come first.” \cite{wachtershauser1992groundworks}

\item “Genetics and metabolism without catalysis is hard to imagine, so proteins must have come first.” \cite{plankensteiner2005prebiotic}

\item “The development of Darwinian selection is hard to imagine without compartments, so membranes must have been there at the outset.” \cite{segre2001lipid}

\end{enumerate}

Ruiz-Mirazo and coauthors \cite{ruiz2013prebiotic} divide the main schools of thought into “gene-first” \cite{joyce1999prospects}, and “metabolism-first” \cite{kauffman2000investigations} and describe metabolism, membrane compartmentation, and genetic replication as “the three main interdependent components of life”. 

It has been cogently argued that self-organized metabolism-like chemical processes (to date, a purely theoretical construct) cannot support the progressive development of molecular complexity,\footnote{In Orgel’s view, “there is no basis in known chemistry for the belief that long sequences of reactions can organize spontaneously---and every reason to believe that they cannot” \cite{orgel1998origin}; Vasas \cite{vasas2010lack} argues for the “lack of evolvability in [hypothetical] self-sustaining autocatalytic networks”.} and that to evolve complexity (metabolic or otherwise) would indeed require a genetic system. To circumvent a cyclic gene/metabolism dependency requires a genetic mechanism that can operate without---yet give rise to---evolved metabol\-ism, preferably in a model that avoids requirements for initial membrane-based compartmentation.

Peptide/IMT models avoid these intractable cyclic dependencies as follows:

\begin{enumerate}

\item Imprinting processes suggest mechanisms by which peptide polymers could \textit{indirectly} template their replicas. If so, 
then genetic systems and Darwinian evolution do not require pre-existing nucleic acids.

\item Prebiotic chemistries and wet-dry cycles can produce amino acids and drive their polymerization without metabolism. Thus, metabolism need not precede the availability of building blocks and energy.

\item Surface imprints can act as selective catalysts and could potentially enable peptide replication through product-templated ligation and chain extension. If so, then enzymatic catalysis need not precede genetic processes.

\item Wet-dry cycling on pore-bearing surfaces can enable both compartmentation and genetic exchange. Thus, membranes need not precede the operation of Darwinian selection on compartmented metabolism.

\end{enumerate}

Further, the generic nature and consequent diversity of amorphous imprinting media, together with the generic occurrence of highly specific interactions between imprints and instances of broad classes of molecules, avoids the requirement for the occurrence of a delicate balance of physical and chemical properties in a specific, self-complementary polymer.

To summarize, peptide/IMT models are broadly compatible with what we know of high-entropy prebiotic chemical environments, do not require implausible chemistries, could potentially support Darwinian evolution of functional complexity, and do not engender intractable cyclic dependencies among necessary developments en route to cellular life. Both experimental and theoretical studies could help to resolve critical open questions in models of this class.

\begin{figure}[tb]
\begin{center}
\includegraphics[width=0.6\linewidth, trim=0.1cm 0.5cm 0.1cm 0.1cm, clip]{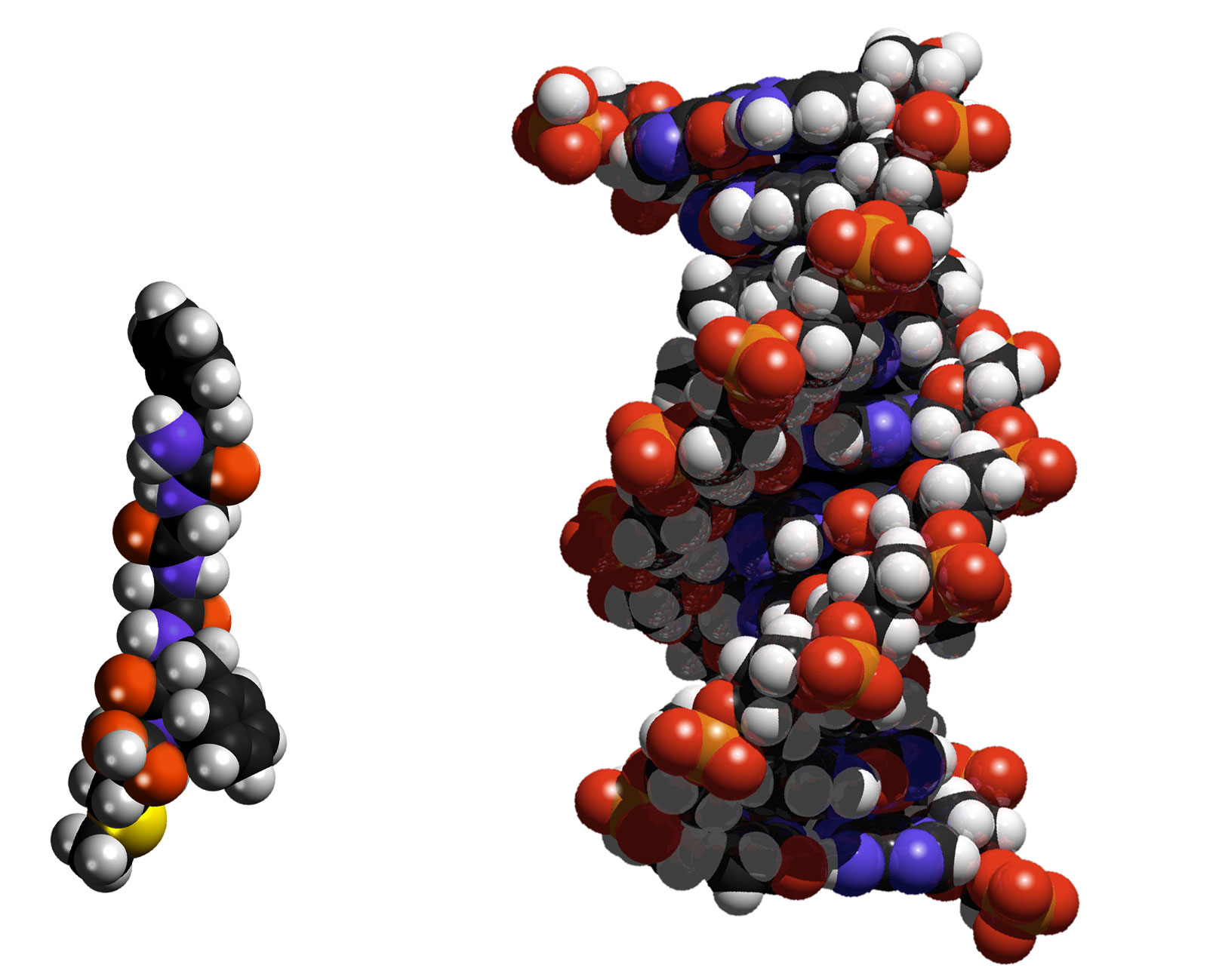}
\caption{A 5-residue peptide (met-enkephalin, left) and 10 base-pair DNA duplex compared. In the context of modern genetic repertoires, the information content of both sequences is $\mathtt{\sim}$20 bits; the DNA:peptide mass ratio is $\mathtt{\sim}$11:1. Amino acids can be produced and peptides can be polymerized (and perhaps replicated) in model prebiotic environments; theoretical considerations and negative experimental results suggest that nucleic acids cannot.}
\label{fig:DNA_v_peptide}
\end{center}
\end{figure}

\section{Concluding summary}
\label{sec:conclusions}

Molecular imprinting is a widely-studied and applied process in which molecules interact with labile amorphous media to produce stable structures (imprints) that exhibit antibody-like binding specific toward molecules of the same kind; these imprints can in some instances act as catalysts for the production of copies of the imprinting molecules.

This paper has extended proposals by Lauterbur for imprint-mediated metabolism \cite{lauterbur2005demystifying} to suggest mechanisms for imprint-mediated replication and elaboration of molecular sequence information in a prebiotic environment, a genetic process that could mark the threshold of cumulative molecular and biological evolution. In contrast to genetic processes based on elusive prebiotic nucleic acids, proposed peptide-based, imprint-mediated genetic processes would:

\begin{itemize}

\item Operate with realistic, racemic, prebiotically-available monomers, rather than requiring complex, homochiral precursors of unexplained origin.

\item Bind and incorporate unactivated monomers from dilute, heterogeneous mixtures of unactivated compounds, rather than requiring separation, concentration, and chemical activation by unknown means.

\item Join monomers to form sequence-specific polymers through product-directed imprint catalysis, rather than requiring pre-existing nucleic-acid polymerase functionality.

\item Directly produce peptides with potentially rich, protein-like chemical functionality, rather than nucleic-acid polymers specialized for inter-strand complementarity at the expense of chemical diversity.

\end{itemize}

Peptide/IMT models share several generic virtues of proposals for surface-based abiogenesis, including reactant concentration and dehydrothermal polymerization though wet-dry cycles, in conjunction with mechanisms for metabolic localization and compartmentation through anchoring and surface porosity. Beyond the threshold of basic genetic functionality, the physical nature of potential peptide/IMT products suggests incremental paths to protein-like structures and complex metabolism, while incidentally explaining the emergence of homochirality through metabolic specialization and consequent symmetry breaking. A direct leap from prebiotic chemistries to nucleic acid replication seems infeasible; a plausible mechanism for the prior evolution of complex metabolism suggests a way in which nature might have bridged this gap.

In brief, the characteristics of proposed imprint-mediated genetic processes suggest solutions to a set of long-standing puzzles. Further investigation may reveal what has so far been elusive: a polymer-mediated genetic templating mechanism that is compatible with realistic prebiotic chemistries and conditions. Pending a truly compelling success in this quest, potential imprint-mediated mechanisms for abiogenesis suggest attractive directions for experimental and theoretical inquiry.

\section*{Acknowledgements}

The author thanks Adam Marblestone, George Church, Gregory Fournier, and Marjorie Cantine for their recent encouragement and suggestions for this paper, and thanks Klaus Mosbach for his hospitality and introduction to the phenomenon of molecular imprinting almost 30 years ago.

\section*{Figure credits}

Images in Figures \ref{fig:nucleotides_v_aminos} and \ref{fig:DNA_v_peptide}, Wikimedia Commons.



\bibliographystyle{alpha}
\bibliography{main}

\end{document}